\documentstyle[preprint,aps]{revtex}
\tightenlines
\begin{document}

\title{Delta function singularities in the Weyl tensor at the brane}

\author{\normalsize{Philip D. Mannheim}\\
\normalsize{Department of Physics,
University of Connecticut, Storrs, CT 06269} \\
\normalsize{Email address: mannheim@uconnvm.uconn.edu}\\
\normalsize{hep-th/0101047, January 8, 2001} \\}

\maketitle

\begin{abstract}
In a recent paper Shiromizu, Maeda and Sasaki derived the gravitational
equations of motion which would hold on a brane which is embedded in a 
higher dimensional bulk spacetime, showing that even when the Einstein 
equations are imposed in the bulk, nonetheless the embedding leads to a
modification of the Einstein equations on the brane. In this comment on 
their work we explicitly identify and evaluate a delta function singularity
effect at the brane which they do not appear to have discussed in their
paper, an effect which, while actually being of interest in and of itself,
nonetheless turns out not to modify their reported results.

\end{abstract}

\bigskip

Interest in the possible existence of extra dimensions has recently been given 
considerable impetus through the work of Randall and Sundrum  
\cite{Randall1999a,Randall1999b} who showed that if our 4-dimensional universe
is a domain wall 3-brane embedded in a 5-dimensional bulk $AdS_5$ spacetime,
the $AdS_5$ bulk geometry would then lead to exponential suppression of the 
geometry away from the brane and thereby localize gravity to it. With
gravitational signals thus effectively being confined to our brane the size
of such extra dimensions could potentially then be very large and yet
nonetheless not lead to conflict with currently available gravitational
information. However, even while such higher dimensions could thus nicely
hide themselves from direct view, it turns out that through the very fact of
there being an embedding into a higher dimensional space at all, the very
presence of extra dimensions then has an indirect effect on gravitational
measurements within our 4-dimensional world, leading (as nicely shown by
Shiromizu, Maeda and Sasaki \cite{Shiromizu1999}) to a modification of the
Einstein equations on the brane, with measurements on the brane potentially
then being able to reveal the existence of higher dimensions and thus probe
their possible existence.\footnote{Such embeddings can even lead to possible
modifications of the equations of state of the fields on the brane, with the
consistency of the embedding often being found \cite{Mannheim2000} to 
lead to negative pressure brane fluids just like the ones currently being
considered in cosmology \cite{Caldwell1998}, except that rather than being due
to the explicit presence of fundamental 4-dimensional quintessence fields,
the negative pressure is instead maintained by gravitational stresses coming
from a higher dimensional bulk.} Thus given the potential significance and 
broad applicability of the work of Shiromizu, Maeda and Sasaki we have
gone over their paper, to find that there is a delta function singularity
effect which they do not appear to have discussed, an effect which, while
actually being of interest in and of itself, nonetheless turns out not to
modify their reported results.

In order to discuss the embedding of our universe into a 5-dimensional bulk
space with metric $g_{AB}$ ($A,B=0,1,2,3,5$) it is particularly convenient
\cite{Shiromizu1999} to base the analysis on the purely geometric Gauss
embedding formula
\begin{equation}
^{(4)}R^{\alpha}_{\phantom{\alpha} \beta \gamma \delta}=
R^A_{\phantom{A}BCD}q_A^{\phantom{A}\alpha}
q^B_{\phantom{B}\beta}q^C_{\phantom{C}\gamma}q^D_{\phantom{D}\delta}-
K^{\alpha}_{\phantom{\alpha} \gamma}K_{\beta \delta}+
K^{\alpha}_{\phantom{\alpha} \delta}K_{\beta \gamma},
\label{1}
\end{equation}   
which relates the 4-dimensional Riemann tensor 
$^{(4)}R^{\alpha}_{\phantom{\alpha} \beta \gamma \delta}$ 
($\alpha,\beta=0,1,2,3$) of a general 4-dimensional surface to the
Riemann tensor $R^A_{\phantom{A}BCD}$ of a 5-dimensional bulk into which
it is embedded via a term quadratic in the extrinsic curvature 
$K_{\mu\nu}= q^{\alpha}_{\phantom{\alpha}\mu}q^{\beta}_{\phantom{\beta}\nu}
n_{\beta ; \alpha}$ of the 4-surface. (Here $q_{AB}=g_{AB}-n_An_B \equiv
q_{\mu\nu}$ is the  metric which is induced on the 4-surface by the embedding
(viz. the one with  which $^{(4)}R^{\alpha}_{\phantom{\alpha} \beta \gamma
\delta}$ is calculated) and $n^{A}$ is the embedding normal.) On introducing
the bulk Weyl tensor $C_{ABCD}=R_{ABCD}-
(g_{AC}R_{BD}-g_{AD}R_{BC}-g_{BC}R_{AD}+g_{BD}R_{AC})/3+
R^E_{\phantom{E}E}(g_{AC}g_{BD}-g_{AD}g_{BC})/12$, contraction of indices in 
Eq. (\ref{1}) immediately allows us to relate the 4- and 5-dimensional
Einstein tensors according to 
\begin{eqnarray}
^{(4)}G_{\mu \nu}=2G_{AB}(q^{A}_{\phantom{A}\mu}q^{B}_{\phantom{B}\nu}+
n^{A}n^{B}q_{\mu\nu})/3-G^{A}_{\phantom{A}A}q_{\mu\nu}/6
\nonumber \\
-KK_{\mu \nu}+K^{\alpha}_{\phantom{\alpha}\mu}K_{\alpha\nu}+
(K^2-K_{\alpha\beta}K^{\alpha\beta})q_{\mu\nu}/2-E_{\mu\nu}
\label{2}
\end{eqnarray}   
where
\begin{equation}
E_{\mu\nu}=C^A_{\phantom{A}BCD}n_{A}n^{C}q^B_{\phantom{B}\mu}
q^D_{\phantom{D}\nu},
\label{3}
\end{equation}   
with the geometric content of Eq. (\ref{2}) being first,
that of the 35 components of $C_{ABCD}$ (viz. the 35 components of the 50
component $R_{ABCD}$ which are independent of $G_{AB}$) 10 of them can be
determined once the induced metric on the 4-surface is known; and second,
that since its left hand side only contains derivatives with respect to the
four coordinates other than the one in the direction of the embedding normal
$n^{A}$, on its right hand side all derivative terms with respect to this
fifth coordinate (labelled $y$ below) must cancel each other
identically.\footnote{For instance, for $ds^2=f(y)(-dt^2+d\bar{x}^2)
+dy^2$, $n^{A}=(0,0,0,0,1)$, term by term Eq. (\ref{2})
yields $^{(4)}G^0_{\phantom{0} 0}=-f^{\prime\prime}/f-f^{\prime
2}/f^2+f^{\prime\prime}/f+f^{\prime 2}/4f^2-f^{\prime 2}/f^2+f^{\prime
2}/4f^2+2f^{\prime 2}/f^2-f^{\prime 2}/2f^2-0$, i.e. $0=0$.}

Dynamical implications of Eq. (\ref{2}) follow on restricting the 
5-dimensional metric to the form $ds^2=q_{\mu\nu}dx^{\mu}dx^{\nu}+dy^2$,
imposing the $y\rightarrow -y$ $Z_2$ Randall-Sundrum brane scenario symmetry
for a 4-surface 3-brane placed at $y=0$ with normal $n^{A}=(0,0,0,0,1)$,
taking the bulk Einstein equations to be of the form
\begin{equation}
G_{AB}=R_{AB}-g_{AB}R^C_{\phantom{C}C}/2=-\kappa^2_5 [-\Lambda_5 g_{AB} 
+T_{\mu \nu}\delta^{\mu}_A\delta^{\nu}_B\delta(y)]
\label{4}
\end{equation}
and imposing the 20 junction conditions 
\begin{equation}
K_{\mu\nu}(y=0^{+})-K_{\mu\nu}(y=0^{-})=-\kappa^2_5(T_{\mu\nu}-
q_{\mu\nu}T^{\alpha}_{\phantom{\alpha}\alpha}/3),~~~
q_{\mu\nu}(y=0^{+})-q_{\mu\nu}(y=0^{-})=0
\label{5}
\end{equation}   
which serve to determine the discontinuity in the extrinsic
curvature at the brane \cite{Israel1966} and enforce the continuity of the
induced metric on it. With the brane symmetry requiring the induced metric
coefficients to be functions of $|y|=y[\theta(y)-\theta(-y)]$ (where
$|y|^{\prime}= \theta(y)-\theta(-y)$, $|y|^{\prime 2}=1$, $|y|^{\prime
\prime}=2\delta(y)$), we see that the extrinsic curvature is an odd,
discontinuous function of $|y|$, while the Einstein and Weyl tensors are even
functions. Consequently, we can determine the extrinsic curvature from the
Israel junction conditions, and since its contribution on the brane is
quadratic in Eq. (\ref{2}) and thus a net even function of $|y|$ which is
continuous across the brane, we can then evaluate its resulting contribution
to Eq. (\ref{2}) by averaging the values of the contributions of these
quadratic terms on the two sides of  the brane. Since the Einstein and Weyl
tensors are even functions of $|y|$ it  is initially very tempting to apply
this same prescription to evaluate their (generically of the form
$\bar{F}=F(y=0^{+})/2+F(y=0^{-})/2$) contributions on  the brane as well, to
thus lead in the $T_{\mu\nu}=-\lambda q_{\mu\nu}+\tau_{\mu\nu}$
case (we conveniently separate out the brane cosmological constant $\lambda$)
to the relation given in \cite{Shiromizu1999}, viz. 
\begin{equation}
^{(4)}G_{\mu \nu}=\Lambda_4q_{\mu \nu}-8\pi G_{N}\tau_{\mu\nu}
-\kappa^4_5\pi_{\mu\nu}-\bar{E}_{\mu\nu}
\label{6}
\end{equation}   
where
\begin{equation}
G_{N}=\lambda\kappa^4_5/48\pi,~~
\Lambda_4=\kappa^2_5(\Lambda_5+\kappa^2_5\lambda^2/6)/2,~~
\label{7}
\end{equation}   
\begin{equation}
\pi_{\mu\nu}=-\tau_{\mu\alpha}\tau_{\nu}^{\phantom{\nu}\alpha}/4
+\tau^{\alpha}_{\phantom{\alpha}\alpha}\tau_{\mu \nu}/12
+q_{\mu\nu}\tau_{\alpha\beta}\tau^{\alpha\beta}/8
-q_{\mu\nu}(\tau^{\alpha}_{\phantom{\alpha}\alpha})^2/24.
\label{8}
\end{equation}   
As we thus see, in the event that the Einstein equations hold in some higher
dimensional bulk spacetime (and even one such as a 10-dimensional one for
instance), in general they will not in fact hold on any lower dimensional
embedded brane as well, with the non-vanishing of the quantities
$\pi_{\mu\nu}$ and $\bar{E}_{\mu\nu}$ signaling an explicit departure from
the standard Einstein equations of motion on the lower dimensional brane. 

Before immediately identifying Eq. (\ref{6}) as the explicit consequence of 
the embedding however, we note that as well as being even functions of $|y|$,
the Einstein and Weyl tensors are also second derivative functions of the
metric. Consequently they must also contain explicit (and equally even)
discontinuous delta function terms in $y$ as well at the brane (cf.
$d^2 f(|y|)/dy^2=d^2 f(|y|)/d|y|^2+2(df(|y|)/d|y|)\delta(y)$ for any  function
$f(|y|)$), terms which then contribute on the brane in addition to the
continuous $\bar{F}$ type averaging contributions. And indeed, the $\delta(y)$
contribution of the bulk Einstein tensor terms in Eq. (\ref{2}) is readily
determined from the Einstein equations, to yield a contribution
\begin{equation}
2G_{AB}(q^{A}_{\phantom{A}\mu}q^{B}_{\phantom{B}\nu}+
n^{A}n^{B}q_{\mu\nu})/3-G^{A}_{\phantom{A}A}q_{\mu\nu}/6=
-2\kappa^2_5[\tau_{\alpha \beta}q^{\alpha}_{\phantom{\alpha}\mu}
q^{\beta}_{\phantom{\beta}\nu}-\tau^{\alpha}_{\phantom{\alpha}\alpha}
q_{\mu\nu}/4]\delta(y)/3
\label{9}
\end{equation}   
on the brane. Now since $^{(4)}G_{\mu \nu}$ is evaluated from the induced 
metric alone, and since $q_{\mu \nu}$ itself is continuous at the brane 
according to Eq. (\ref{5}), $^{(4)}G_{\mu \nu}$ cannot contain any second 
derivatives of $|y|$. Consequently, with Eq. (\ref{2}) being 
a geometric identity, the discontinuous second order derivative terms in
the Einstein tensor must be canceled identically by the discontinuous second
order derivative terms in $E_{\mu \nu}$ (the $K_{\mu \nu}$ dependent terms
are only first derivative functions of $|y|$). Thus as well as having an
average contribution on the brane, $E_{\mu\nu}$ must contain a specific
additional discontinuous $\delta(y)$ dependent term as well,
viz.  
\begin{equation}
E^{disc}_{\mu\nu}=
-2\kappa^2_5[\tau_{\alpha \beta}q^{\alpha}_{\phantom{\alpha}\mu}
q^{\beta}_{\phantom{\beta}\nu}-\tau^{\alpha}_{\phantom{\alpha}\alpha}
q_{\mu\nu}/4]\delta(y)/3,
\label{10}
\end{equation}   
a term which reduces to 
\begin{equation}
E^{disc}_{\mu\nu}=-2\kappa^2_5(\rho_m+p_m)[U_{\mu}U_{\nu}
+q_{\mu\nu}/4]\delta(y)/3
\label{11}
\end{equation}   
for a perfect fluid $\tau_{\mu \nu}=(\rho_{m}+p_m)U_{\mu}U_{\nu}
+p_mq_{\mu\nu}$. Now, as we had already noted above, knowledge of the Einstein
tensor on the brane provides us with information regarding the Weyl
tensor. Equation (\ref{11}) (a relation which is, as far as we know, new) thus
emerges as the dynamical consequence of the absence of any discontinuity in
$^{(4)}G_{\mu\nu}$ or $q_{\mu \nu}$.\footnote{The delta function singularity 
in  $E^{disc}_{\mu\nu}$ arises explicitly  because of the singular nature of
the $Z_2$ symmetric Randall-Sundrum brane set up, with the Weyl tensor on the
brane itself (viz. the one explicitly associated with the induced metric
$q_{\mu\nu}$) not possessing any such singularity.}  Now as far as the
derivation of Eq. (\ref{6}) is concerned, since the delta function terms in
Eqs. (\ref{9}) and (\ref{10}) do cancel identically in Eq. (\ref{2}), only
the average values of the Einstein and Weyl tensors on the brane are
ultimately needed to determine $^{(4)}G_{\mu\nu}$, with Eq. (\ref{6}) as
derived by the averaging procedure thus nicely remaining
intact.\footnote{Bearing in mind the different roles played by even and odd
functions of $|y|$, we note in passing that since the 5-dimensional covariant
conservation condition
$[T^{\mu \nu}\delta^{A}_{\mu}\delta^{B}_{\nu}\delta(y)]_{;B}=0$ which follows
from Eq. (\ref{4}) always involves products of even $\delta(y)$ functions with
odd Christoffel symbol functions whenever $A$ or $B$ is equal to 5, at the 
brane this conservation condition then reduces to the familiar 
$T^{\mu\nu}_{\phantom{\mu \nu};\nu}=0$ (i.e. to $\tau^{\mu\nu}_{\phantom{\mu
\nu};\nu}=0$), a condition which involves the fields and the metric on
the brane alone.}

As regards some possible practical applications of Eqs. (\ref{6}) and
(\ref{11}), we note first that the non-vanishing of $\bar{E}_{\mu \nu}$ 
immediately entails a necessarily non-vanishing Weyl tensor in the bulk and a
thus necessary (and possibly gravity non-localizing
\cite{Mannheim2000,Mannheim2000a}) departure from the maximally 5-symmetric
$AdS_5$ bulk considered in \cite{Randall1999a,Randall1999b}. Second, we note
that even with the vanishing of both $\bar{E}_{\mu \nu}$ and
$E^{disc}_{\mu\nu}$, an $AdS_5$ bulk is still not necessarily secured, since
even if the entire 10-component
$E_{\mu\nu}=C^A_{\phantom{A}BCD}n_{A}n^{C}q^B_{\phantom{B}\mu}
q^D_{\phantom{D}\nu}$ were to vanish, the 25 other components of
the bulk Weyl tensor would still not be constrained. A specific case in point
is the embedding of the $\rho_m=0$, $p_m=0$ (and thus $\rho_m+p_m=0$) exterior
Schwarzschild metric on a brane with cosmological constant $\lambda$ and
normal $n^{A}=(0,0,0,0,1)$ into a bulk with cosmological constant
$\Lambda_5=-\kappa^2_5\lambda^2/6$, a situation for which there is an
explicit exact solution, one in which every single term in Eq. (\ref{6})
vanishes, viz. \cite{Brecher2000} 
\begin{equation}
ds^2=e^{-2|y|}[-(1-2MG/r)dt^2+dr^2/(1-2MG/r)+r^2d\Omega]+dy^2,
\label{12}
\end{equation}   
a solution in which the bulk Weyl tensor components $C_{0101}$
(=$2MGe^{-2|y|}/r^3$), $C_{0202}$, $C_{0303}$, $C_{1212}$, $C_{1313}$ and 
$C_{2323}$ explicitly do not vanish away from the  brane, with the bulk thus
not being $AdS_5$ in this particular case. 

Now while we have identified $E^{disc}_{\mu\nu}$ as the discontinuous piece
of $E_{\mu \nu}$, it is important to note that since $d^2(\sum
a_n|y|^n)/dy^2=\sum a_n n(n-1)|y|^{n-2}+2a_1\delta(y)$, it is thus
possible for $E^{disc}_{\mu\nu}$ ($\sim 2a_1\delta(y)$) to be non-vanishing
even when
$\bar{E}_{\mu\nu}$ ($\sim 2a_2+6a_3|y|+...$) itself does vanish, with the
entire $E_{\mu\nu}$ then being just a delta function at the brane. In fact we
have even found a case in which this explicitly occurs, specifically the
embedding into $AdS_5$ of a static, $p_m \neq -\rho_m$ perfect fluid
Roberston-Walker brane with non-zero spatial 3-curvature $k$, a model which
had been studied in \cite{Mannheim2000,Mannheim2000a}. In this specific case
the solution to Eqs. (\ref{4}) and (\ref{5}) was found to be of the
form\footnote{In passing we note that this particular model was studied in 
\cite{Mannheim2000,Mannheim2000a} since it provides an explicit case where
having a non-vanishing cosmological constant $\Lambda_5$ in the bulk proved
not sufficient (cf. the presence of both converging and diverging
exponentials) to localize the geometry to the brane.} 
\begin{equation}
ds^2=-dt^2e^2(y)/f(y)+f(y)[dr^2/(1-kr^2)+r^2d\Omega]+dy^2 
\label{13}
\end{equation}
where 
\begin{eqnarray}
f(y)=\alpha e^{ \nu |y|}+\beta e^{- \nu |y|}-2k/\nu^2,~~
e(y)=\alpha e^{ \nu |y|}-\beta e^{-\nu|y|},~~
\nu=(-2\kappa^2_5\Lambda_5/3)^{1/2},
\nonumber \\
3\nu(\beta-\alpha)=(\alpha+\beta -2k/\nu^2)\kappa^2_5(\lambda+\rho_m),~~
6\nu(\alpha+\beta)=(\alpha-\beta)\kappa^2_5(\rho_m+3p_m-2\lambda),
\label{14}
\end{eqnarray}   
with the vanishing of the Weyl tensor in the bulk (as required if the bulk
is to actually be the maximally 5-symmetric $AdS_5$ despite the embedding into
it of a much lower, only maximally 3-symmetric, Robertson-Walker brane) then
being enforced \cite{Mannheim2000,Mannheim2000a} by the additional requirement
that $\alpha\beta-k^2/\nu^4=0$, with the fields then having to obey 
$\kappa^2_5(\rho_m+\lambda)(2\rho_m+3p_m -\lambda)=6\Lambda_5$. However,
since for the metric of Eq. (\ref{13}) 10 components of the bulk Weyl tensor
(the 6 $C_{\mu\nu\mu\nu}$ with $\mu\neq\nu$ and the 4 $C_{\mu 5\mu 5}$) would
not in fact vanish without this additional requirement, and since these
components are all kinematically proportional to
\begin{equation}
C_{0505}=e[2eff^{\prime \prime}-3ef^{\prime 2}+2efk+
3e^{\prime}ff^{\prime}-2f^2e^{\prime \prime}]/4f^3,
\label{15}
\end{equation}   
then even with the imposition of the $\alpha\beta-k^2/\nu^4=0$ condition, we
see that while these 10 Weyl tensor components (and thus also the averaged
$\bar{E}_{\mu\nu}$) would then vanish everywhere else, they will still
possess non-vanishing delta function singularities ($C_{0505}\simeq
k\delta(y)$) at the brane.\footnote{While the continuous piece of
$e^{\prime\prime}/e- f^{\prime \prime}/f$ is cancelled by the other terms in
Eq. (\ref{15}), its delta function term
($=\kappa^2_5(\rho_m+p_m)\delta(y)=-4k\delta(y)/[\nu(\alpha -\beta)]$) is
not, with the resulting geometry thus having the somewhat bizarre structure
of being conformal to flat both in the bulk and on the brane while having a
Weyl tensor which nonetheless is not everywhere zero.} In this particular
model then the non-vanishing of $E^{disc}_{\mu\nu}$ is not associated with the
non-vanishing of the Weyl tensor anywhere else, and thus even while a
non-vanishing $\bar{E}_{\mu\nu}$ would immediately signal a non
$AdS_5$ bulk, in and of itself a non-vanishing $E^{disc}_{\mu\nu}$ does
not.\footnote{With manipulation of Eq. (\ref{6}) leading
to \cite{Shiromizu1999} $6\bar{E}^{\mu\nu}_{\phantom{\mu
\nu};\nu}+\kappa^4_5(\rho_m+p_m)(U^{\mu}U^{\nu}+q^{\mu
\nu})(\rho_m)_{;\nu}=0$, we note in passing that for a perfect fluid the
only way that $E^{disc}_{\mu\nu}$ could not be zero even when
$\bar{E}_{\mu\nu}$ does vanish is when $p_m +\rho_m$ is non-zero and $\rho_m$
is spatially homogeneous, this intriguingly being none other than the
situation which prevails in the standard Robertson-Walker cosmology.} 

The author would like to thank Drs. A. H. Guth and A. Nayeri for some very 
helpful discussions. The author would also like to thank Drs. R. L. Jaffe and
A. H. Guth for the kind hospitality of the Center  for Theoretical Physics at
the Massachusetts Institute of Technology where part of this work was
performed. This work has been supported in part by funds provided by the U.S.
Department of Energy (D.O.E.) under cooperative research agreement 
\#DF-FC02-94ER40818 and in part by grant \#DE-FG02-92ER40716.00.


\begin{thebibliography}{99}

\bibitem{Randall1999a} L. Randall and R. Sundrum, 
Phys. Rev. Lett. {\bf 83}, 3370 (1999).

\bibitem{Randall1999b} L. Randall and R. Sundrum, 
Phys. Rev. Lett. {\bf 83}, 4690 (1999).

\bibitem{Shiromizu1999} T. Shiromizu, K. Maeda and M. Sasaki,
Phys. Rev. D {\bf 62}, 024012 (2000).

\bibitem{Mannheim2000} P. D. Mannheim, Phys. Rev. D {\bf 63}, 024018 (2001).

\bibitem{Caldwell1998} R. R. Caldwell, R. Dave and P. J. Steinhardt, Phys. 
Rev. Lett. {\bf 80}, 1582 (1998).

\bibitem{Israel1966} W. Israel, Nouvo Cim. B {\bf 44}, 1 (1966);
{\bf 48}, 463(E) (1967).

\bibitem{Mannheim2000a} P. D. Mannheim, Constraints on $AdS_5$ Embeddings, 
MIT-CTP-2989, hep-th/0009065.

\bibitem{Brecher2000} D. Brecher and M. J. Perry, Nucl. Phys. B {\bf 566},  
151 (2000). 

\end{thebibliography}
\end{document}